\documentstyle[12pt]{article}
\begin{document}

\title{The Generalized Uncertainty Principle ,
entropy bounds and black hole (non-)evaporation in a thermal bath}

\author{Paulo S\'ergio Cust\'odio and J.E.Horvath\\
\it Instituto de Astronomia, Geof\'\i sica e Ci\^encias Atmosf\'ericas\\
Rua do Mat\~ao, 1226, 05508-900 S\~ao Paulo SP, Brazil\\ Email:
foton@astro.iag.usp.br}

\maketitle

\abstract{We apply the Generalized Uncertainty Principle
(GUP) to the problem of maximum entropy and evaporation/absorption
of energy of black holes near the Planck scale. We find within
this general approach corrections to the maximum entropy, and
indications for quenching of the evaporation because not only the
evaporation term goes to a finite limit, but also because
absorption of quanta seems to help the balance for black holes in
a thermal bath. Then, residual masses around the Planck scale may
be the final outcome of primordial black hole evaporation.}

\bigskip
\section{Introduction}
\bigskip

There has been a great deal of interest in the Holographic
Principle [1] (the generalization of the Bekenstein limit
[2]) as a new and fundamental principle in physics, leading
important issues and clues to the unified theories of elementary
fields and forces [3,4].

This limit is very important for black hole physics (as discussed,
for example, in Ref.3), information theory, and several
other fields. There are different derivations of the Bekenstein
bound and the Holographic Principle, but we shall recall here the
simplest intuitive approach.

Let us consider a region with size $R$ containing total energy
$E$. Due to the uncertainty relation the minimum energy
$\epsilon(R)$ of a particle localized inside this region is
$\epsilon(R)\sim{1/R}$ (we use natural units, $L_{pl}\, = c \, = 
\hbar = k_{B} = 1$, however, we restore occasionally $L_{pl}$ in important 
formulae for the sake of clarity). The value $\epsilon(R)$ can 
be considered as a quantum of
energy for that 3-D region. Therefore, the maximum number of
particles inside this region with bounded energy $E$ could be
estimated as the ratio

\begin{equation}
N_{max}\sim{E\over{\epsilon(R)}}\sim{ER}
\end{equation}

Since the Boltzmann entropy of this system is

\begin{equation}
S \, = \, \log \, {\Omega(N)}
\end{equation}

then a maximal entropy of this system is obtained if we consider
that the number of microstates is given by $\Omega(N)={2}^{N}$ (for 
a simple system in which each degree of freedom has just two states 
and no degeneration of the levels). But if the number of
particles (or quanta) is bounded, then we obtain an upper limit to
the entropy $S_{max}< \alpha{ER}$, where $\alpha$ is a calculable
number that accounts for the lack of detailed information of the
specific system under consideration. The presence of a finite
number of species and the internal degrees of freedom do not
change this upper bound.

\section{The Generalized Uncertainty Principle}

\bigskip

Heisenberg obtained the Uncertainty Principle on very general
grounds, using only the quantization of the electromagnetic
radiation radiation field. The uncertainty in the position of the
electron (when it interacts with a photon) is given by
$\Delta{x_{H}}\Delta{P}\sim{1}$, in natural units. He did not,
however, consider gravitational interaction between the electron
and the photon, as is usually assumed to be negligible. However,
at increasingly large energies this interaction is more and more
important.

Here we follow the same arguments of Ref. 3 to derive a
Generalized Uncertainty Principle, later applied to the
generalization of the Bekenstein limit. The fate of extremely
small black holes is discussed within this approach to show an
apparent tendency of remnants with masses around the Planck mass
$M_{pl}$ to remain in equilibrium with a radiation bath.

The simplest way of thinking of this gravitational uncertainty is
to follow a Newtonian approach. If we suppose that the photon is a
classical particle with energy $E$ and effective mass
$m=E/{c}^{2}$, then this particle will interact with the electron
imparting to it an acceleration given by $a = Gm/{r}^{2}$. We
consider that the interaction region has size $L$ and the
interaction lasts $c\Delta{t}\sim{L}\sim{r}$. Then, the
variation of the velocity of the electron is given by
$\Delta{v}\sim{GE\over{c^{2}r^{2}}}{c\Delta{t}\over{c}}
\sim{GE\over{c^{3}L}}$. Then, the position uncertainty due to this
interaction may be obtained by inverting
${v}\sim{\Delta{x}\over{\Delta{t}}}\sim{GE/c^{3}L}$, then
$\Delta{x_{G}}\sim{GEc\Delta{t}\over{c^{4}L}}\sim{GP\over{c^{3}}}$.
The impulse given to the electron is $P\sim{\Delta{P}}$.
Therefore, the total uncertainty in the position is

\begin{equation}
\Delta{x}=\Delta{x_{H}}+\Delta{x_{G}}={1\over{\Delta{P}}}+{L_{pl}}^{2}\Delta{P}
\end{equation}

This relation is invariant under
$\Delta{P}\rightarrow{1\over{\Delta{P}}}$. However, it is unclear
whether we can simply add up the contributions linearly as done
above. The reasons of concern arise from the fact that the
gravitational and the electromagnetic fields interact
non-trivially at these energy scales, and may produce terms 
$\propto \, \sqrt{\Delta{x_{H}}\Delta{x_{G}}}$ Then, an
alternative suggestion may be

\begin{equation}
\Delta{x}=\Delta{x_{H}}+\Delta{x_{G}}+\sqrt{\Delta{x_{H}}\Delta{x_{G}}}
\end{equation}

which is also invariant under the transformation $\Delta{P}\rightarrow{1\over{\Delta{P}}}$.

A General Relativistic approach, free of the action-at-distance
Newtonian drawbacks may be attempted. The field equations
of General Relativity are

\begin{equation}
G_{\mu\nu}={8\pi{G}\over{c^{4}}}T_{\mu\nu}
\end{equation}

The left side has $(length)^{-2}$ units. Thus, on dimensional grounds we may write the
left side in terms of deviations of a flat metric, as

\begin{equation}
LHS\sim{\delta{g_{\mu\nu}}\over{L^{2}}}
\end{equation}

and $\delta{g_{\mu\nu}}$ denotes this deviation and $L$ is a
lengthscale. The energy-momentum tensor $T_{\mu\nu}$ is roughly
the photon energy by $L^{-3}$. Thus, we may write

\begin{equation}
R.H.S \, \sim \,
{8\pi{G}\over{c^{4}}}{E\over{L^{3}}}\sim{GP\over{c^{3}L^{3}}}
\end{equation}

and finally, equating both dimensional estimates yields

\begin{equation}
\delta{g_{\mu\nu}}\sim{GP\over{c^{3}L}}
\end{equation}

This deviation corresponds to an uncertainty in $\Delta{x}$ inside
the region of size $L$. Thus, the uncertainty in the position of
the particle due to this gravitational interaction must be given
by

\begin{equation}
{\Delta{x_{G}}\over{L}}\sim{\delta{g_{\mu\nu}}}\sim{{GP\over{c^{3}L}}}
\end{equation}

And it implies $\Delta{x_{G}}\sim{{G\Delta{P}\over{c^{3}}}}$. This
is the same relation obtained in the Newtonian approach.

String theory provides yet another way of deriving this result,
the details can be found in Refs. 4. The GUP arises on
quite general grounds and may be applied to re-derive known
results, but also to physical systems for which a detailed theory
does not give definitive results, either because not yet completed
or because of the difficulty of the calculations.

\subsection{The Holographic Bound near the Planck scale}

\bigskip

One of these examples is the issue of a maximum entropy within a
bounded region of space, currently subject to a great deal of
activity [2]. To obtain the generalization of the
Holographic Bound we start with eq.(3) and consider a finite
region of size $R$ filled with quanta with minimum energy given by
$c\Delta{P}\sim{\epsilon(R)}$. The GUP expression leads
immediately to

\begin{equation}
{{\epsilon}^{2}(R)}-R{\epsilon(R)}+1\leq{0}
\end{equation} 

Solving for the minimum

\begin{equation}
{\epsilon(R)}_{min}={R\over{2}}\biggl[1-\sqrt{1-{(2L_{pl}/R)^{2}}}\biggr]
\end{equation}

Then, as before, the maximum number of quanta within the region with size 
$R$ and total energy $E$ is

\begin{equation}
N_{max}={E\over{\epsilon_{min}}}={2E\over{R}}{\biggl[1-\sqrt{1-{(2L_{pl}/R)^{2}}}\biggr]}^{-1}
\end{equation}

The number of microstates is $2^{N_{max}}$, and the maximum Boltzmann entropy just

\begin{equation}
S_{max}={\alpha}{2E\over{R}}{\biggl[1-\sqrt{1-{(2L_{pl}/R)^{2}}}\biggr]}^{-1}
\end{equation}

In the limit $R >> L_{pl}$ when the region is much larger than the
Planck length $L_{pl}$ , $S_{max}={\alpha}ER$ as expected (which follows 
from $\epsilon_{min}\approx 1/R$), but
there is a deviation from the linear behavior for small $R$. It must 
be kept in mind that the
ratio ${S_{max}\over{E}}$ can be related to the number of bits
that cross the unit area. Note that the square brackets requires
strictly that $R > 2 L_{pl}$ in order to keep the entropy real,
then, around the Planck scale we found a minimal value for the
Bekenstein bound. Then, we may say that the use of the GUP
excludes values below $\sim 2L_{pl}$ naturally. The spacetime
would be quite fuzzy at this scale, and we can not obtain
information from areas below $\sim 4{L_{pl}}^{2}$. Fig. 1 show the
behavior of ${S_{max}\over{E}}$ versus the size of the system
under consideration $R$ illustrating this feature. If an
alternative expression for the GUP, such as eq.(4) is used the
lower bound on $R$ becomes $3 L_{pl}$, which may be equally
acceptable. This feature may be also inferred from the proposed form 
of the GUP relations eqs. (3) and (4) reminding that the spread 
in the momentum $\Delta P$ must be real. It must be noted that 
there is, in fact, no {\it a \, priori} reason to discard the choice 
of the ''+'' sign for the square root in eqs. (14) and (15); however 
as Adler, Chen and Santiago pointed out [5], that choice 
would not reproduce the Hawking temperature in the limit of large 
black holes, a result that has been derive using a number of 
physical arguments. Therefore, we have also chosen to keep 
the ''-'' sign to be consistent with that limit.

\section{GUP effects on the evaporation/absorption of black holes}

\bigskip

It is well-known that evaporating black holes emit particles with
a thermal spectrum characterizing the Hawking temperature
$T_{H}={M_{pl}^{2}\over{8\pi{M}}}$. This effect can be attributed
to the vacuum polarization induced by the gravitational collapse
and holds for several kinds of black holes (including proper
modifications due to rotation, charge, etc). It is important to
recall that the Hawking temperature above was deduced by applying
the principles of Quantum Mechanics to General Relativity. This
effect is a semiclassical approximation considering quantum fields
around a collapsing body described classically by the General
Relativity. If we modify the Heisenberg Uncertainty Principle as
above, we expect that the Hawking temperature will be modified,
since the simple formula $T_{H}\propto{M}^{-1}$ is a reflection of
the uncertainty relating the size $\Delta{x}$ and the wavelength
${\Delta{\lambda}^{-1}}$ of the field fluctuations.

Adler,Chen and Santiago (Ref.5 and references therein)
analysis used the GUP to derive a modified black hole temperature
exactly as above. From the eq.(10), we solve for the momentum
uncertainty in terms of the distance uncertainty, which we again
take to be the Schwarzschild radius $R_{g}$. This gives the
following momentum and temperature for the radiated photons (and
other particles)

\begin{equation}
\Delta{P}\sim{\Delta{x}\over{2L_{pl}^{2}}}\biggl[1-\sqrt{1-{({2L_{pl}\over{\Delta{x}}})}^{2}}\biggr]
\end{equation}

And therefore, according to their analysis, the corrected formula
for the temperature is

\begin{equation}
T_{GUP}={M\over{4\pi}}\biggl[1-\sqrt{1-{({M_{pl}\over{M}})}^{2}}\biggr]
\end{equation}

which behaves as $T_{GUP}\rightarrow{T_{H}}$ in the limit $M \,
\gg \, M_{pl}$. As already noted, the uncertainty relations 
eqs.(3) or (4) also admit the solution with a "+" sign in front 
of the square root, but this choice would not become the 
standard Hawking temperature in the large mass limit. Other 
physical reason for discarding this possibility is not known to us.

Now, it is important to recall that radiation incident onto a
small black hole can be absorbed by the object if the temperature
of the environment $T_{rad}$ is high compared to the black hole
temperature $T_{H}$ (in agreement with the second law of
thermodynamics). Then the absorption condition is $T_{rad}(t) >
T_{H}(t)$. Geometrically, we know that it happens if the
wavelength of the incident radiation is smaller than the
gravitational radius of the black hole. In the very early
universe, inside the radiation-dominated era, the density evolves
as $\varrho_{rad}(t)\sim{8\times{10}^{5}{(t/s)}^{-2}g{cm}^{-3}}$
assuming zero curvature, i.e. an universe which was flat at its
very early epochs.

The geometric condition $\lambda_{max} < R_{g}$ for absorption of
radiation quanta onto the black hole then becomes a familiar
thermodynamic condition if we substitute Wien's law and the
Hawking temperature in these formula
(($\lambda_{max}T_{rad}=constant$ and $R_{g}T_{H}=constant$)),
because we obtain $T_{rad} > T_{H}$, as naively expected.

Then, all black holes immersed in a radiation environment are able
to accrete this ambient radiation provided the geometric condition
is met. When the geometric condition is not valid, the environment
is colder than the black hole, and the evaporation dominates over
accretion. The evaporation/accretion regimes of black holes have
been studied in the semiclassical approach, and while it has been
shown that a large room for the accretion regime exists, this
regime does not lead to explosive growing when a radiation "fuel"
is considered. Hawking [6] showed that an evaporating black
hole must loss mass at a rate given by

\begin{equation}
{dM\over{dt}}=-{A(M)\over{M^{2}}}
\end{equation}

This result is obtained considering that the energy flux is
thermal and the Stefan-Boltzmann law valid, i.e.
$-\dot{E}=4\pi{R_{g}}^{2}{\sigma}{T_{H}}^{4}$. Numerically,
$A(M)\sim{10}^{24}g^{3}s^{-1}J(M)$, where $3<J(M)<100$ denotes the
particles emitted by the black hole. Their emission probability is
sizeable when $T_{H} > m$, where $m$ is the rest mass of the given
particle. On the other hand, the classical cross section for
absorption is

\begin{equation}
\sigma_{r_{g}}={27\pi\over{4}}{R_{g}}^{2}
\end{equation}

And the mass accretion is given by the product
$\sigma_{r_{g}}F_{rad}(T)$, where $F_{rad}(T)=c\varrho_{rad}(T)$
is the flux of radiation around the event horizon of the black
hole (non-trivial effects due to the full quantum-mechanical 
treatment of the problem
for large black holes are described by N.Sanchez in Ref.7
and references therein).

Therefore, the total variation of mass would be given by

\begin{equation}
{dM\over{dt}}=-{A\over{M^{2}}}+B{M}^{2}{T_{rad}}^{4}(t)
\end{equation}

where $B={27\pi\over{M_{pl}^{4}}}$ in natural units. It is very
interesting to study the equilibrium conditions for these objects
at given epochs. In general, if the mass rate is null, the black
hole is in thermodynamical equilibrium with its surroundings, and
therefore considering $\dot{M}=0$ above, we arrive at the {\it
critical mass} defined by

\begin{equation}
M_{c}(t) \sim {10^{26} \, g \over{T_{rad}(t)/T_{0}}}
\end{equation}

where $T_{0}$ is the present temperature of the CMBR . Today,
$M_{c} \sim 10^{-7} \, M_{\odot}$ and all isolated stellar black
holes are actually absorbing the meager relic radiation from the
background. Assuming that the cosmological mass density in black
holes is subdominant (see Refs. 8 and 9 and references
therein), the time dependence of the critical mass is quite
simple. It has been further shown that the equilibrium between
black holes and radiation is not stable, and therefore the black
holes will not remain in equilibrium after the condition $M
\sim{M_{c}}$ is met. Large black holes ($M \gg M_{pl}$) would
drift away the critical mass curve to the evaporating regime.

What happens with black holes near the Planck scale? When
considering the same problem near the Planck scale, we may improve
our equations taking into account the GUP to study the behavior of
small primordial black holes formed in the very early universe.
The quantum evaporation term may be constructed using the
Stefan-Boltzmann law again
$-\dot{E}=4\pi{R_{g}}^{2}{\sigma}{T_{GUP}}^{4}$ to obtain

\begin{equation}
-\biggl({dM\over{dt}}\biggr)_{evap}={{\sigma}{M}^{6}\over{16{\pi}^{3}M_{pl}^{4}}}
{\biggl(1-{\sqrt{1-{M_{pl}^{2}\over{M^{2}}}}}\biggr)}^{4}
\end{equation}

In order to deduce the absorption term for these very small PBHs,
it is enough require that the critical mass to be the same in this
regime. The argument for this criterion is that the critical mass
is a parameter describing the properties of the bath, not of a
black hole in particular [8,9]. The critical mass rather
describes the equilibrium between a reservoir and any black hole
with generic mass, therefore it has to be an invariant function,
in spite that the absorption and emission terms might vary.
Furthermore, its value is such that, even though derived for
semiclassical conditions, gives a value $\sim M_{pl}$ when the
temperature increases towards the Planck scale, therefore it seems
to comprise information about the near-Planckian regime as well.
We write the absorption term in the form

\begin{equation}
\dot{M}_{abs}={27\pi{M}^{2}\over{M_{pl}^{4}}}F_{rad}(T) \Xi(M)
\end{equation}

and the total variation as (after defining
$\Delta=\sqrt{1-{(M_{pl}/M)}^{2}}$)

\begin{equation}
\biggl({dM\over{dt}}\biggr)=-{{\sigma}{M}^{6}\over{16{\pi}^{3}M_{pl}^{4}}}
{\bigl(1-\Delta\bigr)}^{4}
  +{27\pi{M}^{2}\over{M_{pl}^{4}}}F_{rad}(T) \Xi(M)
\end{equation}

Requiring the critical mass to be the function given in eq.(19),
the function $\Xi$ is found 

\begin{equation}
\Xi(M)={(1-\Delta)}^{4}{\biggl({M\over{\sqrt{2}M_{pl}}}\biggr)}^{8}.
\end{equation}

The interesting property of such changes in the
emission/absorption properties is that all the higher-order
derivatives $\ddot{M},...$ vanish on the critical curve. This
suggests that the mass remains unchanged once it reaches the
near-Planckian regime, stabilizing itself at a value $\sim
M_{pl}$. Thus, the semiclassical evaporation extended down to 
the near-Planckian regime gives a calculation framework to 
implement the quantum feature of non-evaporating black holes 
of size $\leq 2-3 L_{pl}$. 
Black holes would not evaporate completely, but leave a
microscopic residue which, depending on formation scenarios, may
even be cosmologically relevant .

\section{Conclusions}

\bigskip
Let us summarize the main results of this work. First, we have
shown that at very small scales (close to Planck scale),
non-trivial corrections due to the GUP are introduced to the
Holographic Bound to the entropy. In particular, we have show that
the entropy bounds hold for physical systems bigger than $2-3
L_{pl}$. Loop quantum gravity [10] recently yielded a
consistent result for a quantized minimal area, namely $4 \log 3
\times L_{pl}^{2}$. It is tempting to suggest a link between these
results coming from different approaches.

Second, the GUP has been applied in order to understand the
equilibrium of very small PBHs in a thermal bath. The Hawking
evaporation must be changed to the new regime, and therefore, the
time scale for complete evaporation is altered [5]. But we
argued that we must also describe how energy absorption is
modified, since the event horizon is subject to a position
uncertainty according to the GUP. The effective cross-section for
absorption rises close to the Planck scale, as dictated by the
critical mass. Thus, thermodynamical equilibrium conditions are
changed. For extremely small PBHs, the total evaporation is
delayed as a consequence of the reduced emission rate induced by
the fuzziness of the horizon, as well as the contribution of the
modified absorption of hard quanta from the environment. Since the
entropy bound reaches a minimum just above Planck scale, the
absorption must be bounded (whereas emission certainly is,see [10]), 
and suggests the quenching of the evaporation due to
quantum Planck-scale physics. We have argued that absorption is
important to describe the process, since an approach containing
the emission term only is necessarily incomplete for the very
early universe. While based on general arguments only, the 
results support earlier claims about a residual mass
from black hole evaporation [11], an issue which merits
further studies.

\section{Acknowledgements}

Both authors wish to thank the S\~ao Paulo State Agency FAPESP for
financial support through grants and fellowships. J.E.H. has been
partially supported by CNPq (Brazil). We would like to thank the
scientific advice from E. Abdalla on several aspects of this work.

\section{References}

1) R.Bousso, hep-th/0203101 (2002).

2) J.Bekenstein, {\it Phys. Rev. D}{\bf 7}, 2333 (1973).

3) P. Chen and R.J. Adler, gr-qc/0205106 (2002).

4) G. Veneziano, {\it Europhys. Lett.}{\bf 2}, 199 (1986); E.
Witten, {\it Phys.Today}, (1996).

5) R.J. Adler, P. Chen and D. I. Santiago, {\it Gen. Rel.
Grav}{\bf 33}, 2101 (2001).

6) S.W.Hawking, {\it Comm.Math.Phys.}{\bf 43}, 199 (1974).

7) N. S\'anchez, gr-qc/0106222 (2001).

8) P.S.Custodio and J.E.Horvath, {\it Phys.Rev.D}{\bf 58}, 023504
(1998)

9) J.D.Barrow, E.J. Copeland and A.R. Liddle, {\it Phys. Rev.
D}{\bf 46}, 645 (1992); see also {\it MNRAS} {\bf 253}, 675
(1991).

10) O. Dreyer, gr-qc/0211076 (2002); L. Motl, gr-qc/0212096 (2002)
and references therein

11) S. Alexeyev et al. , gr-qc/0201069 (2002).

\vfill\eject
\bigskip
Figure caption

\bigskip

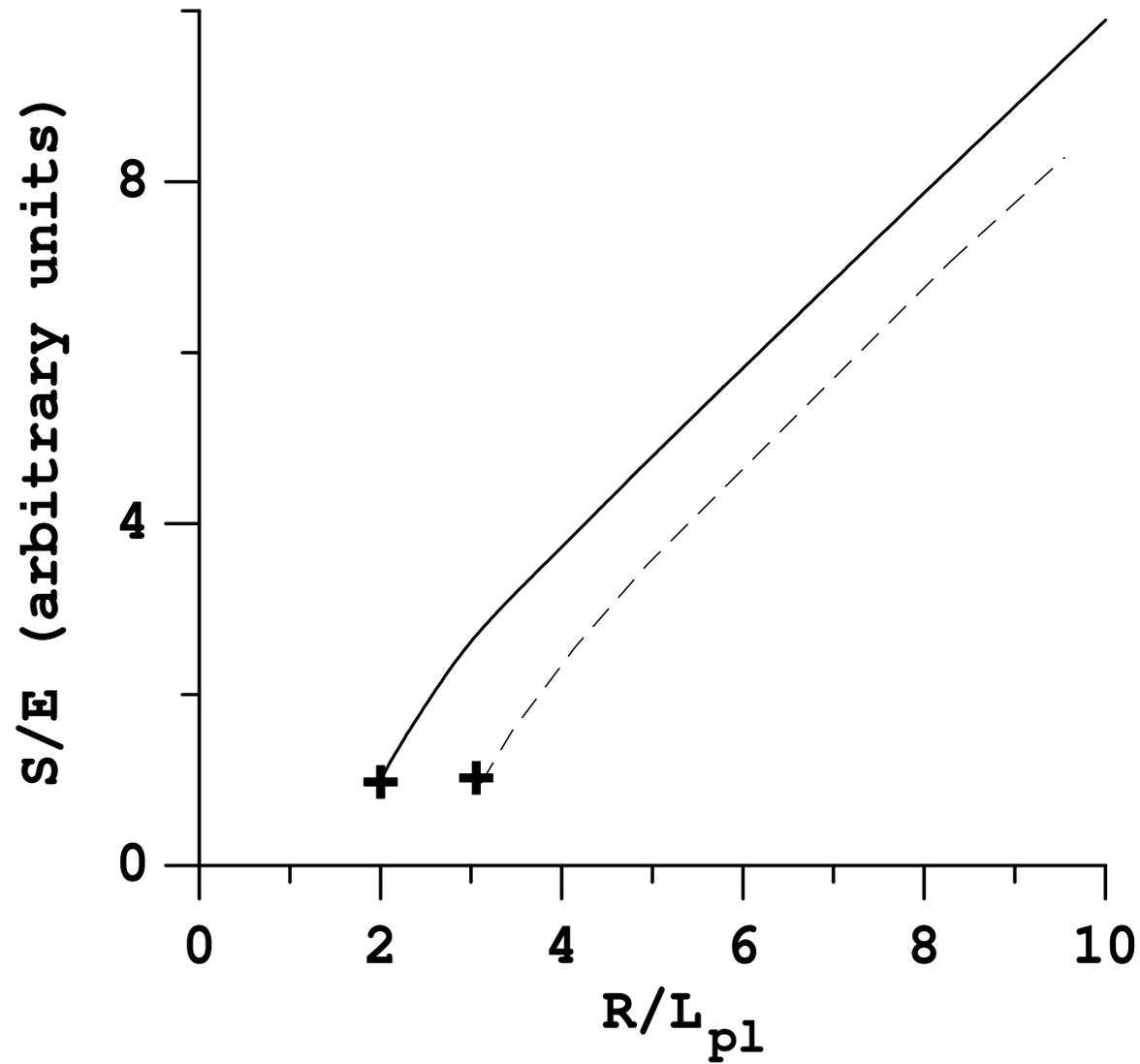
Figure 1.  The Holographic Bound to the entropy for near-Planckian
systems. The amount of information crossing the unit area is
proportional to the vertical axis quantity $S/E$, plotted against
the physical size of the system under consideration $R$. Note that
a real value of $S$ is obtained for $R \geq 2 L_{pl}$ (or $3
L_{pl}$ depending on the exact form of addition of the
uncertainties, eqs.(3) and (4). The decreasing amount of
information allowed may ultimately limit the accretion onto
extremely small black holes (last term of eq.18)

\end{document}